\documentclass[twocolumn]{aastex61}
\usepackage{graphicx}
\usepackage{multirow}
\usepackage{epstopdf}

\usepackage{color}
\usepackage{soul}

\shorttitle{Environmental dependence at high redshifts}
\shortauthors{Gu et al.}

\begin{document}
\title{The effect of environment on star formation activity and morphology at $0.5 < z < 2.5$ in CANDELS}

\correspondingauthor{Qirong Yuan}
\email{yuanqirong@njnu.edu.cn}

\author{Yizhou Gu}
\affil{School of Physics Science and Technology, Nanjing Normal University,
Nanjing  210023, China;
yuanqirong@njnu.edu.cn}
\affil{Department of Astronomy, School of Physics and Astronomy, and Shanghai Key Laboratory for Particle Physics and Cosmology, Shanghai Jiao Tong University,
Shanghai 200240, People’s Republic of China}

\author{Guanwen Fang}
\altaffiliation{Guanwen Fang and Yizhou Gu contributed equally to this work}
\affil{School of Mathematics and Physics, Anqing Normal University, Anqing 246011, People’s Republic of China; wen@mail.ustc.edu.cn}

\author{Qirong Yuan}
\affil{School of Physics Science and Technology, Nanjing Normal University,
Nanjing  210023, China;
yuanqirong@njnu.edu.cn}

\author{Shiying Lu}
\affil{School of Astronomy and Space Science, Nanjing University, Nanjing 210093, People’s Republic of China}

\author{Shuang Liu}
\affil{School of Physics Science and Technology, Nanjing Normal University,
Nanjing  210023, China;
yuanqirong@njnu.edu.cn}

\begin{abstract}
To explore the effect of environment on star-formation and morphological transformation of high-redshift galaxies,
we present a robust estimation of localized galaxy overdensity using a density estimator within the Bayesian probability framework.
The maps of environmental overdensity at $0.5< z< 2.5$ are constructed  for the five CANDELS fields.
In general, the quiescent fraction increases with overdensity and stellar mass. Stellar mass dominates the star formation quenching for massive galaxies, while environmental quenching tends to be more effective for the low-mass galaxies at $0.5< z < 1$. For the most massive galaxies ($M_* > 10^{10.8} M_{\odot}$), the effect of environmental quenching is still significant up to $z \sim 2.5$.
No significant environmental dependence is found in the distributions of S\'{e}rsic index and effective radius for SFGs and QGs separately. The primary role of environment might be to control the quiescent fraction. And the morphological parameters are primarily connected with star formation status. The similarity in the trends of quiescent fraction and S\'{e}rsic index along with stellar mass indicates that morphological transformation is accompanied with star formation quenching.

\end{abstract}

\section{Introduction} \label{sec:intro}
It is well known that the galaxies in the universe can be broadly subdivided into two classes: quiescent galaxies (QGs) with spheroidal morphologies and few star formation, and star-forming galaxies (SFGs) characterized by disc-like morphologies and intensive star formation \citep{Strateva01, Baldry04}. The previous studies have demonstrated that star formation activities and morphologies of galaxies are closely correlated out to $z\sim 2.5$ (e.g., \citealp{Strateva01, Ball06, Franx08, 2011ApJ...742...96W, 2017MNRAS.471.2687B, Gu18}).
Star formation activities and morphologies of the observed galaxies are probably the results involved by several complicated physical processes, such as violent merge, tidal harassment, disk instability, gas cycle, and energy feedback from active galactic nucleus (AGN) or starburst \citep{Conselice14, Somerville15}.  In general, two major factors (internal and external) are considered to play the crucial roles in the evolution of galaxies, which can change the star forming statue of galaxies from star forming to quiescent and maybe change the appearance we observed meanwhile \citep{Baldry06, Peng10, Darvish16, Kawinwanichakij17}. However, the physical mechanisms behind the processes of star formation quenching and morphological transformation, and how these processes change with environment and redshift, are still on debate.

The mechanisms driven by internal physical processes are often referred to as ``mass quenching'' \citep{Peng10}. The energy feedbacks from active galactic nuclei and supernovae are regarded as the internal mechanisms which cease the star formation in galaxies by heating, expelling, and consuming gas \citep{Larson74, Croton06}. Massive bulge in the center of a galaxy may help to maintain the stability of gas dynamics in the disk to avoid collapsing,  which is referred to as “morphological quenching” \citep{Martig09}.
The slow rearrangement of energy and mass results in the internal secular evolution driving gas migration from the outskirts to the center, and suppresses star formation in a long timescale \citep{KK04}.

It is known that the universe is web-like, which are composed of clusters, galaxy filaments, great walls, and large voids on large scale. Besides the internal physics, it has been long established that the external environment in which galaxies reside is another crucial factor for galaxy evolution. The galaxy environment in the local universe is supposed to influence the galaxy properties, such as colours \citep{Balogh04, Blanton05, Bamford09}, star formation rates (SFR) \citep{Kauffmann04, Peng10}, and morphologies \citep{Dressler80, Goto03, Skibba09}. Generally speaking, the galaxies residing in dense  environments tend to be older, redder, more spheroidal and less star forming. Various physical mechanisms are commonly invoked to explain the effects driven by the galaxy environment. For example, ``strangulation'' refers to the mechanism that the cease of gas supply leads to exhaust the remaining gas in a long timescale, and finally to be quiescent \citep{Larson80, vandenBosch08, Peng15}. 
Ram pressure stripping can strip the cold gas rapidly  and result in the suppression of star formation in a short timescale due to the interactions between galaxy and intra-cluster/group media \citep{G&G72}. Similarly, the cumulative effect of many weaker encounters takes gas away from a galaxy by tidal forces, which is referred to as ``galaxy harassment'' \citep{F&S81}. Mergers and strong galaxy-galaxy interactions are also the assignable triggers to shape the galaxy properties fundamentally, including star formation, angular momentum, morphology and nuclear activity \citep{Toomre&Toomre72, Hopkins08}.



In the local universe, star formation activities and morphologies of galaxies are closely related with their environments (e.g., \citealp{Goto03, Baldry06, 2007ApJ...670..206V, 2008MNRAS.383..907B}).  
 Based on the Galaxy Zoo project, it is found that  the morphology-environment correlation is weak at a given color, but color  still strongly depends on environment at fixed morphology \citep{Skibba09}; And at fixed stellar mass, color is more sensitive to the variation of environment than morphology \citep{Bamford09}. They conclude that these is an excess of environment dependence for color compared with that for morphology. \cite{2017MNRAS.471.2687B} also report that the morphologies of massive galaxies are strongly correlated with specific star formation rates (SSFRs) and independent with the environments, which indicates that the local massive galaxies are dominated by the physical process shaping the morphology and determining the star-forming state at the same time.

Using the data from DEEP2 galaxy survey, \cite{2007MNRAS.376.1445C} suggest that the color-density relation is the consequence of environment effects at $z \lesssim 1.3$.
From zCOSMOS galaxy redshift survey, the morphology-density relation persists up to $z \sim 1$ at fixed stellar mass, and becomes flatter towards the high-mass end \citep{2009A&A...503..379T}. For a give Hubble type of galaxies at z = $0.4 – 0.8$, no significant evidence is found that star formation (traced by the [OII] equivalent width) depends on local environment \citep{2008ApJ...684..888P}.
More recently, \cite{Paulino-Afonso19} report a stronger dependence of morphology on stellar mass than environment at $z\sim0.8$.
\cite{2019MNRAS.490.1231L} estimate the time-scale associated with accretion and quenching, and it is long at $0.55 < z< 1.4$, suggesting that the rapid environmental processes (e.g., ram pressure stripping and galaxy harassment) may not be the primary process in this cosmic period.
It is well reported that the separable effects of stellar mass and environment on galaxy properties have been observed at $z \sim 1$ \citep{Baldry06, Peng10, Muzzin12, Darvish16}. It is evidenced that gravity environment has probably played a role in galaxy evolution already at the early epoch. Measuring how the star formation activities and the morphologies of galaxies change with galaxy mass, environmental density, and redshift will undoubtedly help to constrain the environmental effects on galaxy evolution as a function of cosmic epoch.

The high-$z$ clusters provide the ideal laboratories for studing evolution of galaxies within dense environments at earlier time. The studies of the two galaxy clusters at $z > 2$ reveal that the number galaxies are mostly dominated by star-forming systems, which gives a new insight into environmental effect at early epochs \citep{Wang16, Darvish20}. A stronger suppression of star formation is found by \cite{Newman14} under the environment of a cluster at z = 1.8. \cite{Sazonva20} find that the galaxies in two of four clusters possess morphologies distinguishable from the galaxies in fields, but it is not the case for the other two clusters.  The lack of well-defined galaxy clusters at high redshifts raises the potential uncertainty. And the incompleteness of quiescent galaxies haunts the studies of the spectroscopically confirmed clusters or spectroscopic surveys. The discrepant results may be caused by different cosmic epochs, dynamical states of clusters, and statistical bias.

Building up a map of environmental density based on deep surveys opens up another way to investigate environmental effects. Recently, these is an increasing number of studies to quantify the environmental density up to $z \sim 2$.
\cite{Fossati17} reconstruct the density map using the projected density within a fixed aperture of $r_{\rm ap}$ = 0.75 kpc.
\cite{Kawinwanichakij17} introduce the Bayesian estimator by considering the distances to all $N$ nearest neighbors \citep{Ivezic05, Cowan08}.
\cite{Guo17} tactfully define the projected distance between the low-mass galaxy and the nearest massive galaxies as an environmental indicator.
\cite{Ji18} investigate the possible evidence of environmental effects by measuring the small-scale angular correlation function for different types of galaxies.
\cite{Chartab20} present a weighted kernel density estimation by adopting von Mises kernel rather than two-dimension symmetric Gaussian kernel, which is more suitable in the case of the spherical coordinates (e.g, \citealt{Darvish15}).
The different definitions of environment should describe intrinsically disparate physical meanings in different physical scales (e.g., \citealt{Haas12, Muldrew12, Etherington&Thomas15}).


To gain a deeper understanding of environmental effects on galaxy evolution since cosmic noon ($z \sim 2$), in this paper, we  apply the Bayesian-based method \citep{Cowan08} with a correction to construct the overdensity maps for the five fields of the Cosmic Assembly Near-IR Deep Extragalactic Legacy Survey (CANDELS; \citealt{Grogin11, Koekemoer11}).
Firstly, we employ the Bayesian metric as the environmental indicator which takes the distances of all the $N$ nearest neighbors into consideration \citep{Ivezic05}, and exhibit a better performance on the density estimation. The overdensity measurements can be affected by the redshift evolution of comoving number density due to the observation limit. To avoid it, we introduce a correction factor on the overdensity estimation which represents the intrinsic correlation between the number density and the Bayesian environmental indicators. We analyse the mass and environmental effects on star formation quenching and  structural transformation. Then, we explore the environmental effects for SFGs and QGs, respectively.

The layout of this paper is as follow. In Section \ref{sec:data&sample}, we review the basic data from the 3D-HST and CANDELS programs.  In Section \ref{sec:method}, we present the method of the overdensity measurements. The results of quiescent fraction and morphology are shown in Section \ref{sec:fQ} and \ref{sec:morphology}. In Section \ref{sec:discussion}, we further discuss the dependence of morphologies on stellar mass and environment for the star-forming galaxies and quiescent galaxies at $0.5<z<2.5$. The conclusion is summed up in Section \ref{sec:conclusion}.
Throughout the paper, we employ the $\Lambda$CDM cosmology with $H_0=70\,{\rm km~s}^{-1}\,{\rm Mpc}^{-1}$, $\Omega_{\rm m}=0.30$, $\Omega_{\Lambda}=0.70$.

\section{Data Description and Sample Selection} \label{sec:data&sample}
\subsection{Data Description}\label{subsec:data}
The 3D-HST and CANDELS programs provide WFC3 and ACS spectroscopy and photometry over $\sim$ 900 arcmin$^2$ in five fields \citep{Grogin11, Koekemoer11, Skelton14, Momcheva16}. It provides precise  grism spectroscopy and photometric redshifts, which makes possible to estimate the local environmental densities over the redshift range $z \sim 0.5-2.5$. Our galaxy sample is taken from the ``best'' redshift catalogs of the five CANDELS fields \citep{Momcheva16}. It merges their grism-based results with the H-band selected catalogs from \cite{Skelton14}, together with the stellar population parameters, rest-frame colors and SFRs. Here we give a brief description.

The “best” redshifts are organized by giving a priority as spectroscopic redshift, grism redshift, and photometric redshift. The mean uncertainty of photometric redshift in the five fields is $\Delta z /(1+z) \approx 0.02$.  The grism redshifts, with the uncertainty $\Delta z /(1+z) \approx 0.003$ in general, are more accurate  than the photometric redshifts.
The optimal choice is the spectroscopic redshift. When the spectroscopic redshift is not available, the secondary choice would be the grism redshift, then the photometric redshift.
Once redshift is well determined, the rest-frame colors are derived from the filter response function and the best-fit template for each individual source with the EAZY code.
Stellar population parameters are determined with the FAST code (\citealt{Kriek09}), assuming exponentially declining star formation histories, solar metallicity, \cite{Calzetti00} dust extinction law and \cite{BC03} stellar population synthesis models with a \cite{Chabrier03} initial mass function.  The e-folding timescale $\log(\tau/{\rm Gyr})$ varies from 7 to 10  in steps of 0.2, and the age $\log(t/{\rm Gyr})$ varies from 7.6 to 10.1 in steps of 0.1. The dust attenuation ($A_V$) is allowed to vary between 0 and 4 in increments of 0.1.

The rest-frame optical morphologies are depicted by the CANDELS WFC3 J- and H-band images for the galaxies at $z<1.5$ and those at $z>1.5$, respectively.
In this work, the S\'{e}rsic index $n$ and the effective radius $r_{\rm e}$ are regarded as the representatives of galaxy morphology.  Morphology measurements are taken from \cite{vdW14}, in which galaxy images are fitted  by assuming a single S\'{e}rsic profile with the GALFIT (\citealt{Peng02}). The constraints are set  to limit the S\'{e}rsic index from 0.2 to 8, the effective radius from 0.3 to 400 pixels, the axis ratio from $10^{-4}$ to 1, the magnitude from 0 to 40 with an additional condition that the absolute difference with the input value from {\tt SExtractor} \citep{Bertin96} should be less than 3 magnitudes. Here, we refer to \cite{vdW12} for the full description.

\subsection{Sample Selection}\label{subsec:sample}
\begin{figure}
\center
\includegraphics[scale=1.0]{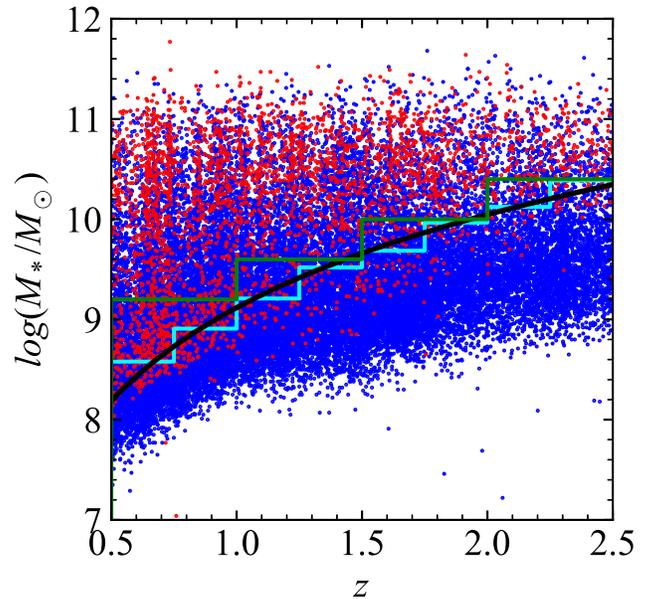}
\caption{Stellar mass as a function of redshift. Passive (red dots) and star-forming galaxies (blue dots) are determined by $U-V$ and $V-J$ colors.
The cyan step-like line shows the mass completeness limits for the UVJ-selected quiescent galaxies in the redshfit interval ($\Delta_z$=0.25).
The mass completeness limits can be parameterized with a function, $M_{\rm comp}(z) = 9.11 + 1.34 \, {\rm ln} (z)$ (black line). The green line step-like line denote the minimums of stellar mass we adopted at the four redshift bins, which are $10^{9.2}$, $10^{9.6}$, $10^{10.0}$, and $10^{10.4}$  $M_\odot$, respectively.
}
\label{fig01}
\end{figure}


We start with a magnitude-limited sample of galaxies. Objects are selected with (1) H-band apparent magnitudes $\rm H_{F160W} < 25$; (2) the redshift range of $0.5 < z < 2.5$; and (3) set the flag {\tt use\_phot} = 1.
Criterion (1) guarantees the uncertainty of photometric redshifts $ \sigma_z = 0.02$,
whereas photometric redshift errors increase to 0.046 at $\rm H_{F160W} = 25 - 26$ \citep{Bezanson16}. The lower limit of criterion (2) is set due to the small volume of the CANDELS survey at lower redshifts, whereas the upper limit is set due to the high $M_{\rm lim}$ at higher redshifts.  Criterion (3) ensures a reasonably uniform quality for the photometry of galaxies. Here we refer to \cite{Skelton14} and \cite{Momcheva16} for the full details.
In our sample, about 30 percent of galaxies have spectroscopic redshifts or grism redshifts. Large sample size and high accuracy of redshift ensure a robust determination of environmental density at $0.5<z<2.5$.

In the magnitude-limited sample, the completeness of stellar mass depends on both the redshift and the mass-to-light ratio $M/L$.  Since quiescent galaxies have higher $M/L$ ratio, the completeness limit of quiescent galaxies is higher than that of star-forming galaxies. To be conservative, we derive the mass limit only using the quiescent population.
We estimate the mass completeness of quiescent galaxies using the method in \cite{Pozzetti10}. The quiescent population is separated from the star-forming population by the $U-V$ and $V-J$ colors selection \citep{Williams09}:
\begin{eqnarray*}
& (U - V) > 1.3, (V - J) < 1.6, \\
& (U - V) > 0.88 (V - J) + 0.59~[0.5 < z < 1.0],  \\
& (U - V) > 0.88 (V - J) + 0.49~[1.0 < z < 2.5].
\end{eqnarray*}
The faintest 20\% of quiescent galaxies are considered to estimate the completeness limit ($M_{\rm comp}$)  with the interval of $\Delta z = 0.25$.
With the typical mass-to-light ratio ($M_*/L$) for each galaxy, the stellar mass limit $M_{\rm lim}$ at a redshift slice can be derived if their apparent magnitude equals to the magnitude limit. In detail, the stellar mass limit $M_{\rm lim}$ at a specified redshift can be derived by
$\log(M_{\rm lim}) = \log(M_*) + 0.4(H-H_{\rm lim})$,
where $H_{\rm lim}$ is set as the magnitude limit of our sample ($H_{\rm F160W}$ = 25).
Then $M_{\rm comp}$ is defined as the upper envelop of the $M_{\rm lim}$ distribution below which lie 90\% of the $M_{\rm lim}$ values at a given redshift. Figure \ref{fig01} show mass completeness limit for the quiescent galaxies at four redshift bins. This method makes sure the 90\% completeness of quiescent galaxies at any redshift. 
The mass completeness limits can be parameterized as a function of redshift, $M_{\rm comp}(z) = 9.11 + 1.34 \, {\rm ln}(z)$, denoted by black line, which describes how the mass completeness limits vary from z = 0.5 to 2.5 \citep{Quadri12}. For analysing the quiescent fractions and morphologies at different redshifts, we define four mass complete subsamples at four redshift bins with the redshift interval of $\Delta z = 0.5$. We only apply 9.2, 9.6, 10.0, and 10.4 as the minimums of stellar mass at the four redshift bins, which have been denoted with the green lines in Figure \ref{fig01}.  We provide the calculated limits of stellar mass completeness for quiescent galaxies and the adopted limits in Table \ref{tab01}.  For our four mass-complete subsamples, the proportion of the galaxies only having photometric redshifts is 37\%, 38\%, 46\% and 53\%. The galaxies with grism and spectroscopic redshifts account for at least half at the four redshift bins. Besides, the proportion of quiescent populations at the four redshift bins are 19\%, 20\%, 28\% and 28\%,  respectively.

\begin{table}[]
\caption{The calculated limits of stellar mass completeness for quiescent galaxies and the adopted limits from z = 0.5 to 2.5. \label{tab01}}
\begin{tabular}{p{1.3cm}p{1.2cm}<{\centering}p{1.25cm}<{\centering}p{0.8cm}<{\centering}p{0.8cm}<{\centering}p{0.8cm}<{\centering}}
\hline \hline
redshift & $M_{\rm comp}$\tablenotemark{a} & $M_{\rm adopt}$\tablenotemark{b}  &  $N$\tablenotemark{c} & $f_{\rm phot}$\tablenotemark{d} & $f_{\rm Q}$\tablenotemark{e} \\ \hline
0.50-0.75 & $10^{8.58}$  & \multirow{2}*{$10^{9.2}$}   &  \multirow{2}*{6196 }     & \multirow{2}*{37\%} & \multirow{2}*{19\%}  \\
0.75-1.00 & $10^{8.91}$         & ~ & ~ & ~ & ~    \\
\hline
1.00-1.25 & $10^{9.21}$         & \multirow{2}*{$10^{9.6}$}   &  \multirow{2}*{4262 }     & \multirow{2}*{38\%} & \multirow{2}*{20\%}  \\
1.25-1.50 & $10^{9.52}$         & ~ & ~ & ~ & ~    \\
\hline
1.50-1.75 & $10^{9.69}$         & \multirow{2}*{$10^{10.0}$}  &  \multirow{2}*{2270 }     & \multirow{2}*{46\%} & \multirow{2}*{28\%}  \\
1.75-2.00 & $10^{9.97}$         & ~ & ~ & ~ & ~    \\
\hline
2.00-2.25 & $10^{10.12}$        & \multirow{2}*{$10^{10.4}$}  &  \multirow{2}*{ 874 }     & \multirow{2}*{53\%} & \multirow{2}*{28\%}  \\
2.25-2.50 & $10^{10.38}$        & ~ & ~ & ~ & ~    \\
\hline
\end{tabular}
\tablenotetext{a}{The completeness limit $M_{\rm comp} (M_\odot)$  at the given redshift. }
\tablenotetext{b}{The adopted limits $M_{\rm adopt} (M_\odot)$  with the redshift interval of $\Delta z = 0.5$.}
\tablenotetext{c}{The corresponding size of mass-complete subsample $N$.}
\tablenotetext{d}{The photometric fraction of subsample $f_{\rm phot}$.}
\tablenotetext{e}{The quiescent fraction of subsample $f_{\rm Q}$.}

\end{table}




\section{The measurement of environmental  overdensities }\label{sec:method}

\begin{figure*}
\center
\includegraphics[scale=0.65]{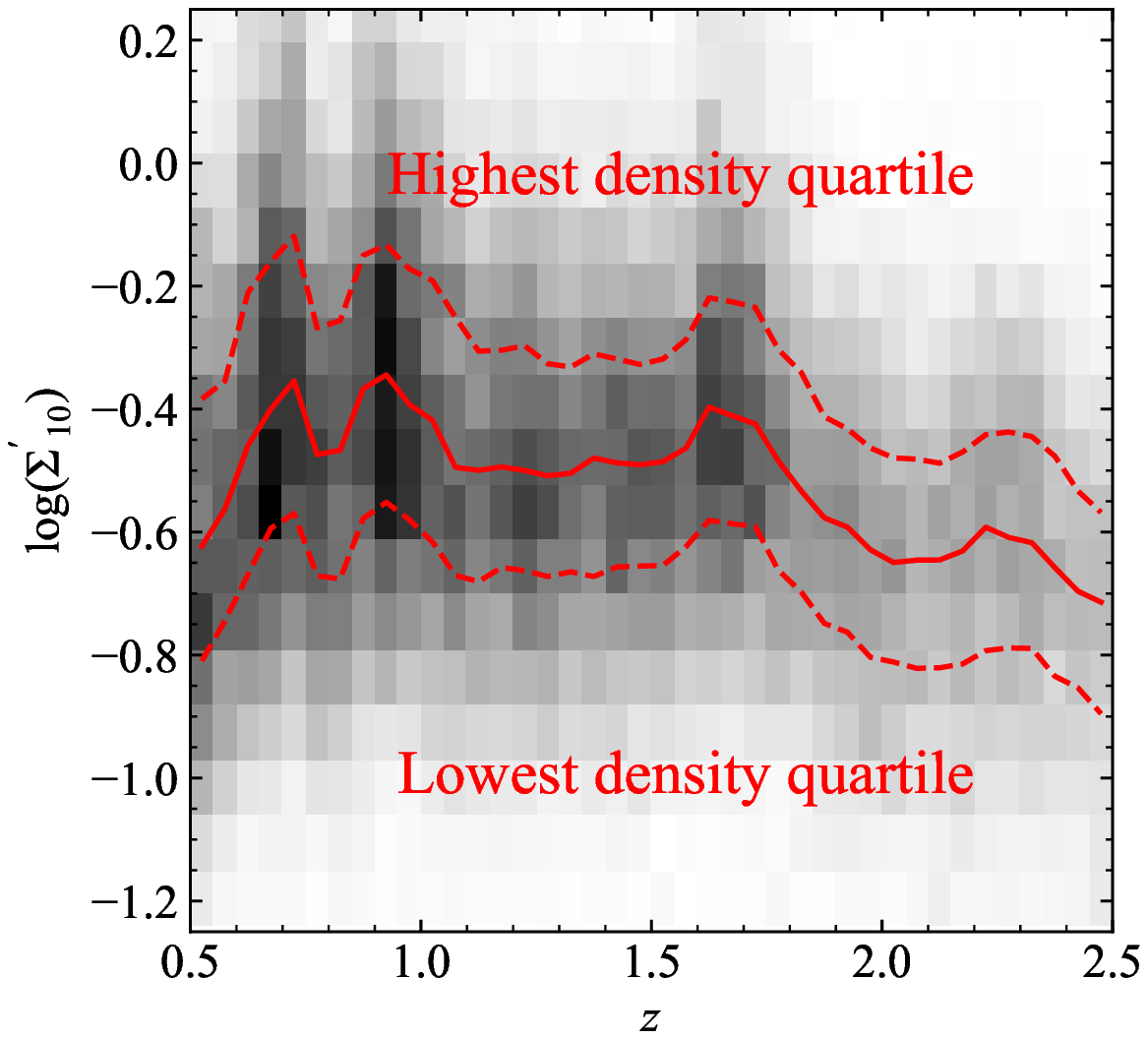}
\includegraphics[scale=0.65]{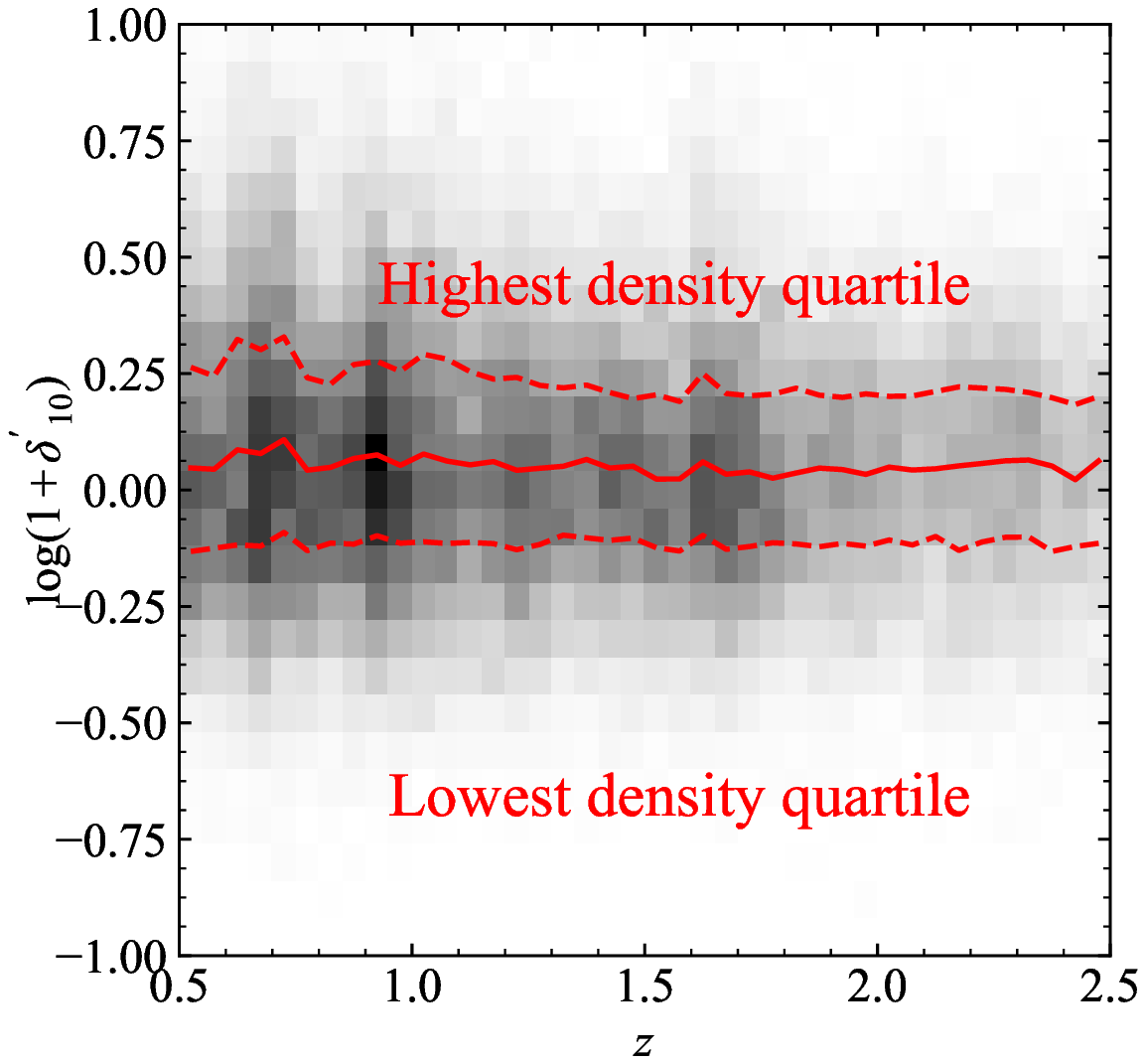}
\caption{Left: Density indicator $\Sigma'_{10}$ as a function of redshift. Right: Overdensity $1 + \delta'_{10}$ as a function of redshift. The red solid and dashed lines are derived by a small redshift interval, $\Delta z = 0.1$, which show the median, bottom, and top quartiles of the distribution. The background represents the distribution of out galaxy sample.
}
\label{fig02}
\end{figure*}

Environmental density can be traced by means of the enclosing neighboring galaxies in the projected plane.
The projected overdensity of galaxies is also referred to as a tracer of local environment.
The nearest-neighbor estimator corresponds to the scale of interhalo environment \citep{Muldrew12}. Although a pure density-based definition of the environment of galaxies (probably due to the view angles) cannot fully separate them into real physical structures, it is still an effective way to trace the most of high-overdensity regions \citep{Shattow13}.
In this work, we adopt Bayesian metric \citep{Cowan08}. It is defined as
\begin{equation}
\Sigma'_N \propto 1/(\Sigma^N_{i=1} d^2_i),
\end{equation}
where $d_i$ is the projected distance towards the $i$th nearest neighbor in projected two-dimensional space of the characteristic redshift bin. The traditional method calculates local environmental density via $\Sigma_N \propto 1/(\pi d^2_N)$, where $d_N$ is the $N$th nearest neighbor distance \citep{Dressler80, Baldry06}. The Bayesian environmental density considers the contribution of all the 1st, 2nd, ... $N$th nearest neighbors which can give the improved accuracy in mapping the probability density distribution compared to the traditional method \citep{Ivezic05, Cowan08}.

Besides the indicator of environmental density, the width of the redshift bin is the other key factor.  Choosing a small width leads to the lack of galaxies which may not reflect the true projected density of a large scale structure. Conversely, a large redshift width would bring about severe contamination in the excess signal of high-density region by foreground and background galaxies.
In this work, the calculation of local overdentsity for a galaxy involves the galaxies within the individual redshift slice of $|\Delta z| = 2\sigma_z(1 + z)$.
Conservatively, we adopt the uncertainty of photometric reshift as the reshift uncertainty ($\sigma_z = 0.02$).
If the probability density function of redshift follows a Gaussian distribution, the setting of 2$\sigma_z$ width ensures that almost members (95.4\%)  in the large scale structure should be taken. But, the contaminants by the randomly superposed galaxies with larger redshift uncertainty may dilute the density estimates to some degree. We also apply a narrower width $\pm 1.5 \sigma_z (1+z)$ as a test, and find no change in the main conclusion.

Due to the observation limit of galaxy survey, the comoving number densities of the observed galaxies tend to be less as the redshift increases. It is foreseeable that the more galaxies are observed inside the volume, the denser the average environmental densities would be, and vice versa.
Thus, we define the dimensionless overdensity, $1 + \delta'_{N}$, as the indicator of galaxy environment:
\begin{equation}
1+\delta'_{N} = \frac{\Sigma'_N}{\langle \Sigma'_N \rangle_{\rm uniform}} = \frac{\Sigma'_{N}}{k'_{N}\Sigma_{\rm surface}},
\end{equation}
where $\Sigma_{\rm surface}$ is the surface number density in unit of arcmin$^{-2}$ within the redshift slice of a given galaxy. The denominator, $\langle \Sigma'_N \rangle_{\rm uniform}$, represents the standard of the Bayesian environmental densities when galaxies distribute in the uniform condition at the given $\Sigma_{\rm surface}$, where $k'_{N}$ is a correction factor which describes the intrinsic linear correlation between $\Sigma_{\rm surface}$ and the standard of the Bayesian density at a given redshift (see the Appendix for detail). In this way, the overdensity becomes more essential in representing the excess of galaxy density. When $\log(1+\delta'_N) > 0$, it means the environmental density for a given galaxy exceeds the density standard in the uniform condition, and vice versa.

The small value of N may cause the fluctuation of the density values due to the Poisson noise and the redshift uncertainty of foreground and background galaxies. In this work, we use Bayesian density estimator based on distances to all 10 nearest neighbors ($\Sigma'_{10}$) as the environmental tracers \citep{Cowan08}.
Figure \ref{fig02} shows that the Bayesian densities $\Sigma'_{10}$ (left) and the overdensities $1 + \delta'_{10}$ (right) as a function of redshift with four quantiles. Since we consider the intrinsic relation between surface number densities and the Bayesian environmental densities, our overdensity represents the value of observed Bayesian density relative to its standard in the uniform condition at a given redshift.
Therefore, the medians and $\pm$25th percentiles of the distribution of overdensities show small amplitude of variation over $0.5<z<2.5$ (see right panel of Figure \ref{fig02}).


\section{Quiescent fraction}\label{sec:fQ}
In this section, we exhibit how the quiescent fraction evolves along with stellar mass and overdensity ($1+\delta'_{10}$) over a wide redshift range of $0.5<z<2.5$.
As shown in Figure \ref{fig01}, the mass completeness limits as a function of redshift are denoted by
black line. To be conservative, the following results are based on the four mass-complete subsamples with the redshift interval $\Delta z = 0.5$.
As discussed in Section \ref{subsec:sample}, we only apply 9.2, 9.6, 10.0, and 10.4 as the minimum logarithms of stellar mass at four redshift bins. The adopted limits of stellar mass, seeing in table \ref{tab01}, are represented with the green lines in Figure \ref{fig01}. 

\subsection{Quiescent fraction as a function of overdensity}\label{subsec:fQ-delta}

\begin{figure*}
\center
\includegraphics[scale=0.5]{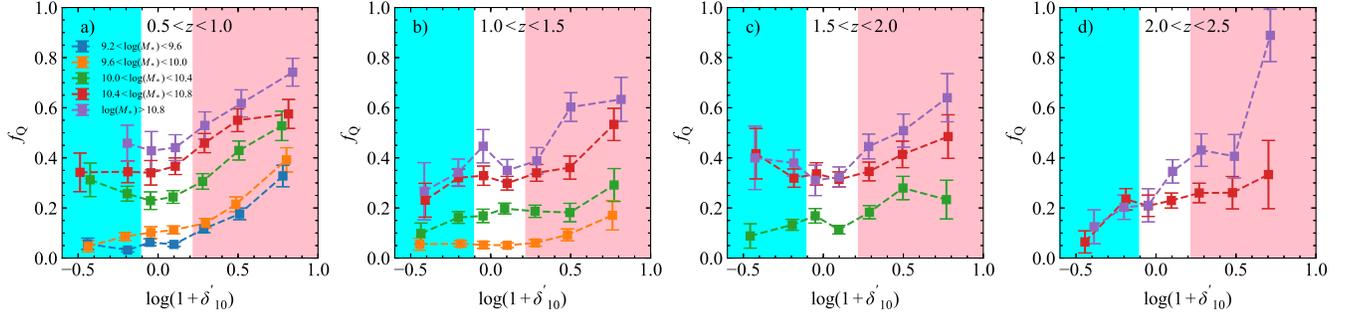}
\caption{The quiescent fractions change as a function of overdensity at a specified stellar mass range in four redshift bins. The different colors represent galaxies in the different stellar mass bins ($\Delta M_* \sim 0.4$ dex). The pink and cyan regions mark the highest density quartile and the lowest density quartile. The error bars indicate the uncertainties based on Poisson statistics.
}
\label{fig03}
\end{figure*}

In this section, we investigate the fraction of quiescent galaxies as a function of overdensity and redshift at fixed mass bins. Figure \ref{fig03} shows the quiescent fractions as a function of the overdensities ($1+\delta'_{10}$) at different stellar mass and redshift bins. Different colors represent different mass bins:
(i) $9.2 < \log(M_*/M_\odot)<9.6$ (blue),
(ii) $9.6 < \log(M_*/M_\odot)<10.0$ (orange),
(iii) $10.0 < \log(M_*/M_\odot)<10.4$ (green),
(iv) $10.4 < \log(M_*/M_\odot)<10.8$ (red), and
(v) $\log(M_*/M_\odot)>10.8$ (purple).
The pink and cyan shaded regions mark the highest quartile and the lowest quartile of the environmental overdensities. The error bars are estimated by the bootstrap method with 1000 times resamplings. 

As shown in Figure \ref{fig03}, the quiescent fractions change as a function of overdensity at the fixed stellar mass bin. At the lowest redshift bin ($0.5<z<1.0$), quiescent fractions of galaxies are susceptible to overdensities for all stellar mass bins. A significantly higher quiescent fraction can be found in denser environments at a fixed stellar mass. The enhancements of quiescent fraction are roughly 20\%$-$40\% as the $\log(1 + \delta'_{10})$ changes from 0 to 0.8.  \cite{Allen16} find that the fraction of quiescent galaxies with  $> 10^{9.5}M_\odot$ increases from 33\% to 55\% from the field to the cluster core at $z \sim 0.95$.  Our results are also in consistent with the previous works \citep{Darvish16, Kawinwanichakij17}.
\cite{Paulino-Afonso18} also find that the less massive galaxies  ($10 < log(M_*/M_\odot) < 10.75$) have a jump of quenched fraction from $\sim 10\%$ to
$\sim 60\%$ at intermediate to higher density regions.
But for $log(M_*/M_\odot) > 10.75$, they find no dependence of the quenched fraction on local density, this being nearly constant at $\sim 30\%$. It may be caused by the different environmental tracers, the different selections of quiescence, and the sampling of spectroscopic observations.

Meanwhile, these are also the differences between quiescent fractions of galaxies with different stellar masses. It reveals that the quiescent fraction at the fixed overdensity tends to be higher for more massive galaxies. In general, the quiescent fraction of galaxies increases with both stellar mass and overdensity at $0.5<z<1.0$.
The limit of out sample is down to $10^{9.2} M_\odot$ at the lowest redshift bin.
Compared with the $\sim 30\%$ enhancements of quiescent fraction by overdensity, we find the difference is not significant in quiescent fraction between $M_* \sim 10^{9.4} M_\odot $ and $M_* \sim 10^{9.8} M_\odot $ at the fixed overdensity. On the other hand, there are clear observable signs of the environmental quenching for the low-mass galaxies ($M_* < 10^{10} M_\odot $) at $0.5 < z < 1.0$.
It implies that the mass quenching may not important for the low-mass galaxies with $M_* < 10^{10} M_\odot$, comparing with environmental quenching.
It supports that environmental quenching plays a more significant role in the truncation of star formation at $0.5 <z < 1.0$ for low mass galaxies, in general agreement with the previous studies (e.g., \citealt{Peng10, 2011MNRAS.411..675S, 2011MNRAS.411.1869L, Darvish16, 2019MNRAS.484.3806L}).

For the galaxies at higher redshift bin ($z>1.0$), their quiescent fraction also changes strongly with overdensity at the highest mass end ($M_* > 10^{10.8} M_\odot $). For the most massive galaxies, it suggests that environmental quenching still plays a role out to $z\sim 2.5$. This is generally consistent with preview works that the influence by dense environments on suppressing star formation still persists up to $z\sim 2.0$,  \citep{Fossati17, Guo17, Kawinwanichakij17, Ji18, Chartab20}.
Interestingly, the quiescent fraction for the highest density bin is unusually high at $2 < z < 2.5$ when compared to all the other samples.
It might be simply caused by the cosmic variance.
However, the ``downsizing'' scenario also suggests that the quenching for the more massive galaxies might accomplish earlier,  in agreement with the quiescent fraction for the highest density bin at $2 < z < 2.5$. For the most massive galaxies,  quiescent fraction is also elevated in the denser environments at all redshift considered. The most massive galaxies in the dense environments seems to have the highest probability to be quenched. It is likely that both the mass and environment quenching mechanisms should have played important roles on the quenching for the most massive galaxies out to $z \sim 2.5$.
Besides this, the overall changes in quiescent fraction from the lowest to the highest overdensities become weaker towards the higher redshift for the galaxies with lower mass. The less massive galaxies at higher redshift bin $z>1.0$ are not so susceptible to the environments in comparison with the corresponding galaxies at $0.5<z<1.0$. We speculate that the galaxy quenching at $z>1.0$ might be dominated by stellar mass (internal processes). In comparison with the galaxies at $0.5 < z < 1$, the environmental quenching has played a less important role at higher redshifts ($1<z<2.5$). Out results in this sections are   qualitatively in agreement with the
previous works \citep{Peng10, 2011MNRAS.411..675S, Muzzin12, 2015ApJ...810...90L, Darvish16, Kawinwanichakij17, Chartab20}.

\subsection{Quiescent fraction as a function of stellar mass}\label{subsec:fQ-lmass}

\begin{figure*}
\center
\includegraphics[scale=0.5]{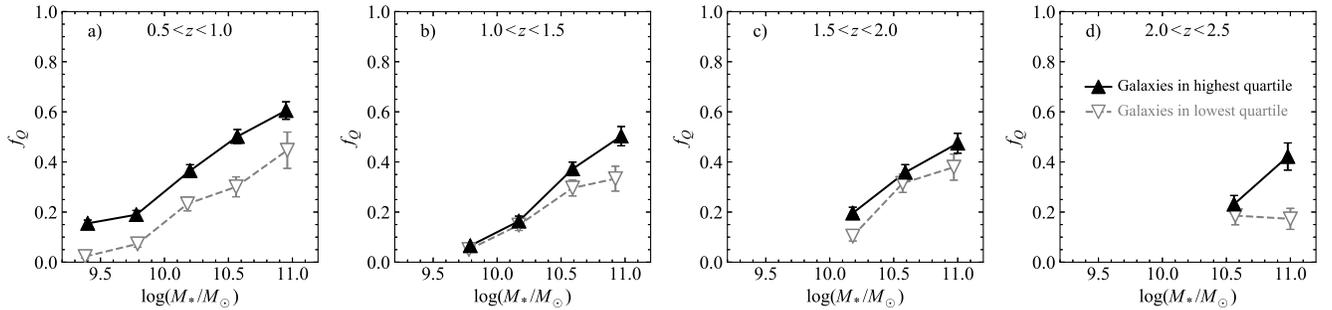}
\caption{Quiescent fraction as a function of stellar mass at fixed overdensity and redshift bins. The hollow and solid symbols correspond to the lowest density quartile and the highest density quartile. The error bars indicate the uncertainties based on Poisson statistics.
}
\label{fig04}
\end{figure*}

It has been shown that the fraction of quiescent galaxies is also a function of stellar mass (e.g., \citealt{Baldry06}). In this section, we investigate how the fraction of quiescent galaxies varies with stellar mass and redshift. Figure \ref{fig04} shows the quiescent fractions of galaxies in the lowest and highest density quartiles as a function of stellar mass at the different redshift bins.  The hollow and solid trangles correspond to the lowest  and the highest overdensity quartiles, respectively.


At a given stellar mass, there is the significant gap ($\sim 20\%$) in $f_{\rm Q}$ between the galaxies in the highest and lowest overdensity quartiles.
Indeed, it supports that the environmental quenching has played an effective role for all the range of stellar mass considered at $0.5<z<1.0$. At $z > 1.0$,  the gap is no statistically significant difference at low-mass end between two overdensity quarters.
However, the fractions of quiescent galaxies are found to be significantly elevated  $\sim 40\%$  from $\sim 10^{9.8} M_\odot$ to $\sim 10^{11} M_\odot $. The median offset of $f_{\rm Q}$ between the two quartiles is roughly half of the promotes by stellar mass.
Noticed that the relation between $f_{\rm Q}$ and stellar mass is weak at the low-mass end (also seeing in the left panel of Figure\ref{fig03}), the environmental quenching tends to be more effective than mass quenching for these low-mass galaxies. According to the results above, we suppose that both mass quenching and environmental quenching are responsibility for the quenching of galaxies at $0.5< z < 1.0$, while the environmental quenching tends to be more effective for low-mass galaxies.

Regardless of  the galaxies in the lowest and highest density quartiles, the mass dependence of $f_Q$ is still established out to $z = 2$. Figure \ref{fig03} also illustrates that the most massive galaxies have quiescent fractions no less than the galaxies with lower mass at a given overdensity in general. This is also in agreement with \cite{Kawinwanichakij17}, who show the increasing fraction of quiescent galaxies with  $> 10^{9.5}M_\odot$ out to $z = 2$.
Similarly, \cite{Darvish16} find that also quiescent fraction depends on stellar mass down to $10^{10}M_\odot$  at $1.5 < z <2.0$.

This mass dependence seems persist only for the galaxies in the highest quantile, but vanish for the galaxies in the lowest quantile at $2.0 < z< 2.5$.
In this work, no significant sign of mass dependence is found in the lowest quantile at $2.0 < z< 2.5$.
However, \cite{Darvish16} find that quiescent fraction in the dense environment depends on stellar mass down to $10^{10}M_\odot$ at $2 < z <3.1$, as well as in the lowest density.
The cause of the discrepancies in the lowest density might be cosmic variance, how the environment is traced, and the different definition of quiescence. The larger contaminants by the galaxies with larger uncertainty of photometric redshift may also tend to dilute the density estimates. The larger and deeper surveys with large spectroscopic and/or photometric data sets are still need to resolve this issue.

It is found that the galaxies in the highest quartile show an increasing trend of quiescent fraction with the raising of stellar mass.  From another perspective, it is above-mentioned that  for the most massive galaxies with $M_* > 10^{10.8} M_\odot $, quiescent fraction is also elevated in the denser environments at all redshift considered. The most massive galaxies in the dense environments seems to have the highest probability to be quenched. The ``downsizing'' scenario points to that the quenching for the most massive galaxies might accomplish in a very short timescale \citep{Cowie96}. It is likely that both the mass and environment quenching mechanisms should have played important roles on  the quenching for the most massive galaxies.
Halo quenching might be an alternative mechanism,
where the massive host halo ($M_{\rm halo}>10^{12} M_\odot$) would heat the infalling cold gas via shocks  (e.g., \citealt{Birnboim&Dekel03}).
The ``overconsumption'' model combines the effects of gas supply disruption (i.e., starvation) and the gas consumption via star formation and outflow \citep{McGee14, Balogh16}, which could also explain the quenching of the most massive galaxies at high redshifts.

\section{Morphologies of galaxies}\label{sec:morphology}
The galaxy morphology is another entry point to investigate the environmental effects on galaxy evolution. The environmental processes have been proposed to play a crucial role on the formation of galaxy morphology \citep{Conselice14, Somerville15}. In this work,  S\'{e}rsic index $n$ and effective radius $r_{\rm e}$  are adopted as the representatives of galaxy morphologies.

\begin{figure*}
\center
\includegraphics[scale=0.5]{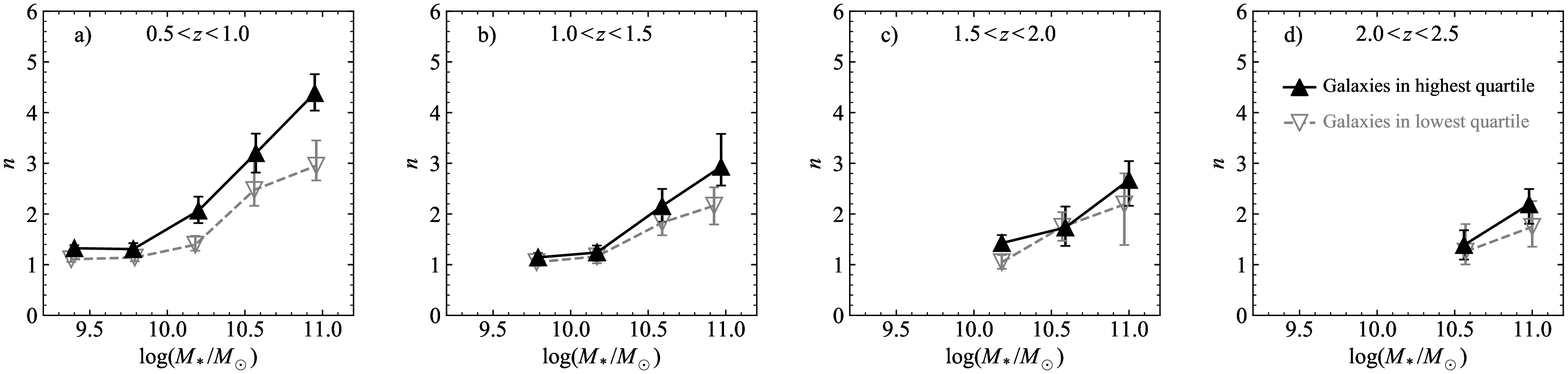}
\includegraphics[scale=0.5]{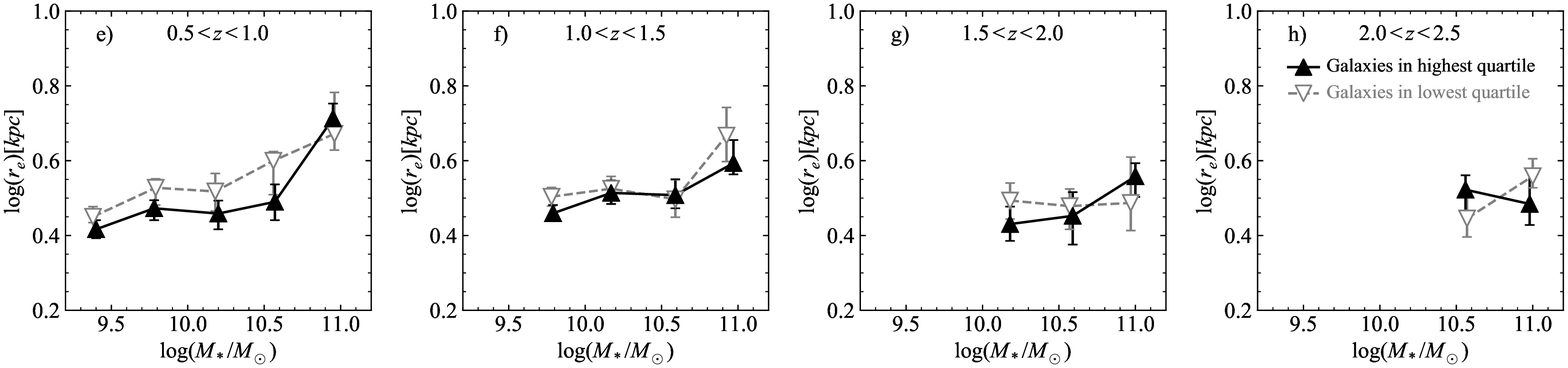}
\caption{Median values of S\'{e}rsic indices and effective radii as a function of stellar mass at fixed overdensity and redshift bins. The hollow and solid symbols correspond to the lowest and the highest overdensity quartiles, respectively.
The errors are estimated by the bootstrap method with 1000 times resamplings.
}
\label{fig05}
\end{figure*}

We aim to investigate how S\'{e}rsic index and effective radius change with stellar mass, overdensity, and redshift. Figure \ref{fig05} shows that the median values of S\'{e}rsic indices and effective radii as functions of stellar mass in different environments. The hollow and solid symbols correspond to the lowest and the highest overdensity quartiles, respectively. The bars represent 1 $\sigma$ errors which are estimated by the bootstrap method with 1000 times resampling.

The variation of quiescent fraction with stellar mass closely resembles the median S\'{e}rsic index as a function of stellar mass.
At the lowest redshift bin ($0.5<z<1.0$), stellar mass dominates the growth of S\'{e}rsic index. As stellar mass increases, the median values of S\'{e}rsic indices in different environments tend to increase in general. It points to that the continuous assembly of stellar component in a galaxy could bring the growth of the central bulge, which leads to gas consumption and star formation quenching. For the low-mass galaxies at $0.5<z<1.0$, however, the quiescent fraction increase faster from the lowest to highest overdensities while the median S\'{e}rsic index remains constant. The change in quiescent fraction is faster than the change in galaxy S\'{e}rsic index towards the low-mass end.
It indicates that some low-mass galaxies quenched by some environmental processes (e.g., ram pressure stripping, strangulation) would retain the morphologies of their star-forming progenitors \citep{Larson80,G&G72}. The massive quiescent galaxies prefer to reside in the high-density environment.
Besides, the divergence between the median values of $n$ for the lowest and highest density quartiles is slight at the low mass end. The divergence  seems to be enlarged towards the high mass end. For our mass-limited subsample at $0.5 < z < 1.0$, the high-density environment can enhance S\'{e}rsic indices of the galaxies at $0.5<z<1.0$, particularly for massive galaxies.

At higher redshifts ($z > 1.0$), the divergences tend to be smaller compared with those at $0.5<z<1.0$. It shows that the environmental influence on S\'{e}rsic index is negligible. As for $z>1$, there is no statistically significant difference (given the error bars) between two overdensity quarters.
On the other hand, there is an evidence that the quiescence of the galaxies still relies on stellar mass at $z > 1.0$. For our mass-limited subsamples, the overall changes driven by mass at $z > 1.0$ are weaker than that at $0.5<z <1.0$.
Therefore, the bulge growth (i.e., increase of S\'{e}rsic index from 1 to 4) is mainly driven by stellar mass. The enhancement of S\'{e}rsic indices by stellar mass is stronger than the environmental influence, even at the lowest redshift bin.

According to the bottom panels of Figure \ref{fig05}, the median size shows the weak dependence on stellar mass at $0.5 < z < 1.0$. For our mass-limited subsample at $0.5 < z < 1.0$, the sizes of the galaxies in dense {\bf environments} are slightly smaller as a whole, except for the most massive galaxies. For our mass-limited subsamples at $z>1$, no obvious evidence of the environmental and mass effect on galaxy size can be found.

\section{discussion}
\label{sec:discussion}

\begin{figure*}
\center
\includegraphics[scale=0.5]{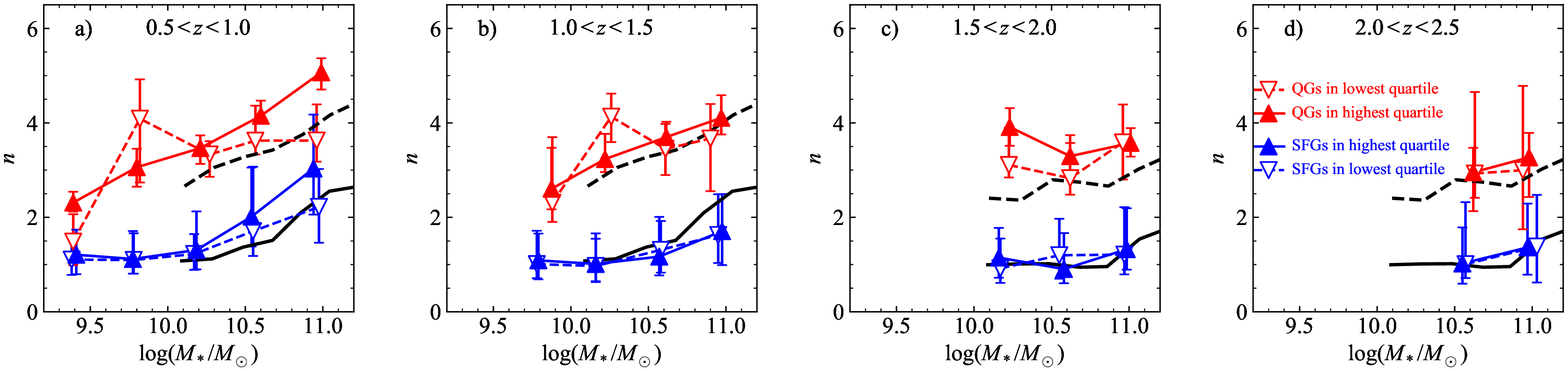}
\includegraphics[scale=0.5]{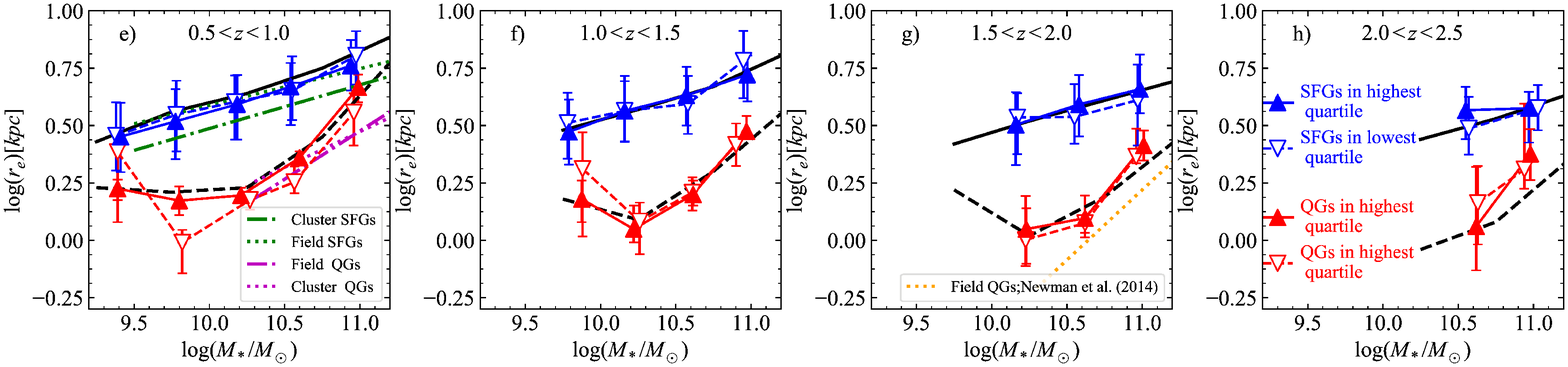}
\caption{Median values of S\'{e}rsic indices and effective radii as a function of stellar mass at fixed overdensity and redshift bins. Blue (Solid) and red (dashed) lines represents star-forming galaxies and quiescent galaxies, respectively. Hollow and solid symbols correspond to the lowest density quartile and the highest density quartile. 
The errors are estimated by the bootstrap method with 1000 times resamplings.
On the top panels, the median values of n as a function of stellar mass from \cite{Lang14} are also presented by black dashed and solid lines, separated by two redshift ranges of  $0.5<z<1.5$ and $1.5 < z < 2.5$.
On the bottom panels, the median values of $r_e$ as a function of stellar mass from \cite{vdW14} are also represented. The size-mass relations from \cite{Allen16} are denoted by green and magenta lines, and a yellow dot line extracted from \cite{Newman14}.
}
\label{fig06}
\end{figure*}

The activity of star formation should be the most direct and intrinsic difference between SFGs and QGs.
\cite{2007MNRAS.376.1445C} conclude that the color-density relation is the consequence of environment effects that the preference of red galaxies in dense environments at $z \lesssim 1.3$. The evidence of environmental quenching also is provided by the decreasing SFR from the outskirt to the center of a distant cluster or from the low overdensity to high overdensity \citep{Muzzin12, Darvish16}.
The SSFR-density trend of a mix sample may reflect the change in the quiescent fraction, rather than the change in SFRs \citep{2009ApJ...705L..67P}.
As more QGs are found in high density region, it is compatible with the results that the SSFR of SFGs is independent of environment at fixed stellar mass\citep{Muzzin12,Darvish16}.
Color might be primary dependent on stellar mass. Once the mass is fixed at $\log(M_*/M_\odot) > 10.7$, the color-density relation is flatten globally \citep{2010A&A...524A...2C}. These relevant works suggests that the primary role of environment is to control the quiescent fraction. Stellar mass might be the primary predictor of star formation.

The morphology-density relation may also be controlled by the change of the ratio between red and blue galaxies.
Similar with the SSFR-density relation, it is found that the morphology-density relation in the local universe implies that the environment can also shape the morphologies of galaxies \citep{Dressler80, Goto03, Bamford09, Skibba09}, which still maintains at higher redshift \citep{2007ApJ...670..206V, 2009A&A...503..379T, 2010ApJ...718...86K, 2013A&A...555A...5N, Allen16, 2017A&A...598A.120K, Paulino-Afonso19}.
By a mean SFR for each morphology types, \cite{2008ApJ...684..888P} predict the SFR-density relation by using the morphology-density relation (the fractions of the morphological types as a function of density). The equivalence of the two relations suggest that neither of two is more fundamental than the other. By assuming the constant value of S\'{e}rsic index (n=1 for SFGs, and n=4 for QGs), \cite{Paulino-Afonso19} predict the overall trends of $n$ depend on stellar mass and environment, which is different in approach but equally satisfactory in result.

Based on the overdensity maps derived by the Bayesian metric for the five CANDELS fields, the large size and wide redshift coverage of our sample facilitate investigating the environmental effect on quiescent faction and morphology. We further explore the environmental effect on galaxy morphologies by splitting galaxies into SFGs and QGs.
In this work, we only compare morphologies of galaxies as a function of stellar mass in two density quarters.

In Figure \ref{fig06}, the median values of S\'{e}rsic indices and effective radii are shown as a function of stellar mass at fixed overdensity and redshift bins. We also overlap the two general relations \citep{Lang14, vdW14}. For the mass-$n$ relation from \cite{Lang14}, the median values of n as a function of stellar mass are also presented at $0.5<z<1.5$ and $1.5 < z < 2.5$.  For the mass-size relvdW14ations from \cite{}, the median values of $r_e$ for QGs and SFGs are denoted by the black dashed and solid lines.
In addition, the size-mass relation for SFGs and QGs \citep{Allen16} are also overlapped by green and magenta lines, with cluster galaxies within 0.5 virial radius and field galaxies denoted by dash-dotted and dot lines. The size-mass relation for QGs at $z = 1.8$ in field from \cite{Newman14} are also presented in yellow dot line.


The environmental influence on galaxy morphologies for SFGs and QGs at high redshifts is still discrepant. \cite{Bassett13} find that quiescent galaxies in the outskirt of a cluster at $z \sim 1.6$ have smaller Sérsic indices, compared to the field galaxies, whereas the SFGs in clusters and fields show no difference in morphology.
However, \cite{Allen16} find that the cluster SFGs within 0.5 virial radius at $z = 0.92$ have a higher fraction of S\'{e}rsic indices with $n > 1$, than field SFGs.
But for QGs, it is consistent in S\'{e}rsic index, $n\sim2$ regardless of the distance from cluster center.
Regardless of the structural difference between SFGs and QGs, \cite{Sazonva20} find that the member galaxies in two of four clusters at $z\sim1.45$ possess structural parameters indistinguishable from the galaxies in fields. But more bulge-dominated galaxies are found in the other two clusters at $z\sim 1.2$ and 1.8, compared with the galaxies in fields. It points out that the morphology–density relationship happens at $z = 1.75$ although there is a significant degree of intracluster variance.
Recently, it is concluded that stellar mass is a stronger predictor of galaxy structure and morphology than local density from the VIMOS Spectroscopic Survey of a Supercluster (${\rm VIS^3COS}$) at $z \approx 0.84$ \citep{Paulino-Afonso19}. They re-products the overal trend of $n$-density relation by $f_Q$-density relation which suggests that the environmental effects mainly control quiescent fraction (e.g., \citealt{Darvish16}).
In this work, no significant mass dependence of $n$ are found at higher redshift ($1.5<z<2.5$).
But for the quiescent galaxies in dense  environments at lower redshifts ($0.5<z<1.5$), it is striking that the median of S\'{e}rsic indices tends to increase with stellar mass. There is only a weak evidence at $0.5<z<1.0$ that massive QGs (or SFGs) in the highest-density environments have larger Sérsic indices, which is also reported by \cite{Kawinwanichakij17}.
Beyond that, there is no statistically significant difference of S\'{e}rsic indices between SFGs (or QGs) in two different environments.

As to the environmental influence on galaxy size, the results are also highly disputed due to the limited numbers of (proto-)clusters and the member galaxies over $z > 1$ \citep{Overzier16}.
It is reported that cluster QGs at $z \sim 1.6$ have larger average effective sizes than field galaxies at fixed mass \citep{Papovich12, Bassett13}. However, \cite{Newman14} find that the quiescent member galaxies at $z=1.8$ follow the size-mass relation in field with a systematic difference $\sim 0.01$.  \cite{Allen16} report that the observed effective radii of SFGs in field are 16\% larger than those in cluster center, but no significant difference for QGs. This comparison between field and cluster galaxies with 0.5 viral radius at $z = 0.92$ is shown the panel (e) of Figure \ref{fig06}. It is also reasonable since the environmental effects on galaxies morphologies may only happen in the center of clusters.
In this work, our work is limited to the two density quarters (the highest and the lowest).
These is only a weak evidence that the effective radii of QGs in highest quartile is larger than those in lowest quartile, compared with the mass dependent of sizes for QGs.
Thus, we conclude that there is no significant evidence exhibiting the difference of effective radii between  QGs (and SFGs) in two different environments.
But the median radius of galaxies in the highest quarter are found slightly smaller than those in the lowest quarter in Figure \ref{fig05}.
In the bottom panels of Figure \ref{fig06}, the different size-mass relations for QGs and SFGs are exhibited that the sizes of SFGs are larger than those of QGs.
Meanwhile, the quiescent fraction of the galaxies in the highest quarter are found slightly larger than those in the lowest quarter.
Considering QGs having smaller sizes systematically, this `trap' should be caused by a higher quiescent fraction in denser environment.




It can be seen that the galaxy morphological parameters ($n$ and $r_{\rm e}$) as functions of stellar mass are distinctly largely different between SFGs and QGs. SFGs usually have larger sizes and are dominated by the disk structures, whereas QGs are bulge-dominated and have smeller sizes (e.g., \citealt{Lang14, vdW14}).
Considering the intrinsic difference in morphology between SFGs and QGs, the overall distributions of S\'{e}rsic indices ($n$) and effective radii ($r_{\rm e}$) should be naturally contributed by two major galaxy populations (SFGs and QGs) at different redshifts.

For QGs and SFGs separately, no significant dependence of morphologies on environmental density is found at $0.5<z<2.5$ compared with the intrinsic difference. In another word, there is no obvious evidence that the morphological differences are caused by environmental processes. It is also an evidence that morphological transformation is accompanied by star formation quenching, which is consist with the results in \cite{Pandya17} and \cite{Gu18}. Thus, we conclude that galaxy morphologies are primarily relevant to the status of star formation quenching. Stellar mass is also an important factor relevant to the sizes of galaxies at $0.5<z<2.5$, which is in agreement with previous works \citep{Lang14, vdW14}. For the S\'{e}rsic indices of galaxies, we only find that the mass dependence at $0.5<z<1.5$.
At $1.5<z<2.5$, S\'{e}rsic indices are not sensitive to stellar mass. There is no obvious evidence that the morphological differences are influenced by some environmental processes. Statistically, quiescent fraction in a sample is conclusive for the distributions of size and S\'{e}rsic index.



\section{Conclusion}\label{sec:conclusion}
In this paper, we present a robust estimation of local galaxy overdensity using an density estimator within the Bayesian probability framework. We build up a map of environmental overdensity in five CANDELS fields across the redshift range $0.5 < z < 2.5$. Then we explore the mass and environmental dependences of quiescent fractions and galaxy structural parameters of the galaxies.

The main conclusions are the followings.

1. Stellar mass is the dominant factor driving star formation quenching of massive galaxies at higher redshifts ($z > 1$) while environmental quenching tends to be more effective for the low-mass galaxies $\log(M_*/M_\odot) < 10.0$ at lower redshifts ($z < 1$).

2. For the most massive galaxies with $\log(M_*/M_\odot) > 10.8$, the effect of environmental quenching is still significant up to $z \sim 2.5$. 


3. It is also found that galaxy morphologies are primarily correlated with star formation status. There is no significant evidence that morphologies are dependent on the environmental density.

4. The distributions of quiescent fractions and S\'{e}rsic indices along with stellar mass is similar in general. We consider that the quiescent fraction is conclusive for the distributions of galaxy sizes and S\'{e}rsic indices. It indicates that morphological transformation is accompanied with star formation quenching.

\acknowledgements
This work is based on observations taken by the 3D-HST Treasury Program (GO 12177 and 12328) with the NASA/ESA HST, which is operated by the Association of Universities for Research in Astronomy, Inc., under NASA contract NAS5-26555.

This work is supported by the National Natural Science Foundation of China (Nos. 11873032, 11673004, 11433005). We acknowledge the science research grants from the China Manned Space Project with NO. CMS-CSST-2021-A07. G.Y.Z acknowledges the support from China Postdoctoral Science Foundation (2020M681281) and Shanghai Post-doctoral Excellence Program (2020218).
G.W.F. acknowledges the support from Chinese Space Station Telescope (CSST) Project, Yunnan young and middle-aged academic and technical leaders reserve talent program (No. 201905C160039) and Yunnan Applied Basic Research Projects (2019FB007).


\appendix
\label{sec:appA}

\section{Derivation of environmental overdensity}
In order to normalize the estimator within the Bayesian probability framework, $\Sigma'_N$, into a dimensionless environment indicator, we define the overdensity as
\begin{equation}
1 + \delta'_N = \Sigma'_N  / \langle \Sigma'_N \rangle_{\rm uniform},
\end{equation}
%
where $\Sigma'_N$ is the Bayesian indicator of environmental density, and $\langle \Sigma'_N \rangle_{\rm uniform}$ represents its standard in the uniform condition. It is important to establish the conversion relation between $\langle \Sigma'_N \rangle_{\rm uniform}$ and the surface number density $\Sigma_{\rm surface}$ for the five CANDELS fields.



\begin{figure}
\center
\includegraphics[scale=0.6]{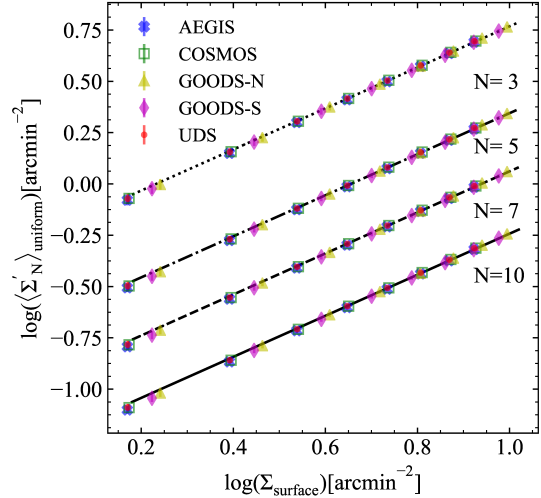}
\caption{The typical value of environmental density indicator in the uniform condition, $\langle \Sigma'_N \rangle_{\rm uniform}$ as a function of surface number density ($\Sigma_{\rm surface}$). The scale factors ($k'_N$) are 0.60, 0.23, 0.12, 0.06 for $N = 3, 5, 7, 10$, respectively.
Tiny uncertainties of $\langle \Sigma'_N \rangle_{\rm uniform}$ are estimated by using 1,000 bootstrap resamples.
The solid lines represent the linear fittings.
}
\label{fig07}
\end{figure}

A simple simulation is made to realize that the galaxies just follow the uniform distribution with a given surface density. For each field of CANDELS, we randomly drop the galaxies into real sky coverage, and the galaxies are uniformly distributed. In practice, we make 8 different uniform maps containing different numbers of galaxies, ranging from 300 to 1700, for each field of CANDELS survey. The Bayesian density estimator, $\Sigma'_N$, is then calculated for each galaxy in the simulated uniform map. The median value of the Bayesian densities $\Sigma'_N$ of all stimulated galaxies can be adopted as the typical value of the overall density indicator in the uniform condition,  $\langle \Sigma'_N \rangle_{\rm uniform}$.
Figure \ref{fig07} shows the relations between the surface number density ($\Sigma_{\rm surface}$) and the typical degree of environmental indicator ($\langle \Sigma'_N \rangle_{\rm uniform}$), with the variable numbers of the nearest neighbors adopted ($N = 3, 5, 7, 10$).
There is a clear trend that typical value of density estimator in the uniform condition is proportional to the given surface number density, i.e. $\langle \Sigma'_N \rangle_{\rm uniform} \propto \Sigma_{\rm surface} $. We introduce a scale factor, $k'_N$, 
and the best-fitting relation
can be simply formalized as:
$\langle \Sigma'_N \rangle_{\rm uniform}$ =  $k'_N \times \Sigma_{\rm surface}$.


Based on above intrinsic linear relation, we propose a new method to estimate the overdensity at a given redshift bin. For each galaxy, we can convert the Bayesian density estimator into the dimensionless overdensity by
\begin{equation}
1 + \delta'_N = \Sigma'_N  / k'_N \Sigma_{\rm surface},
\end{equation}
where $\Sigma_{\rm surface}$ is the surface number density for the given redshift slice.
For each galaxy, we can easily derive the  typical value of the Bayesian density indicator, $\langle \Sigma'_N \rangle_{\rm uniform}$, from the surface number density of the given redshift slice.



\begin{figure*}
\center
\includegraphics[scale=0.72]{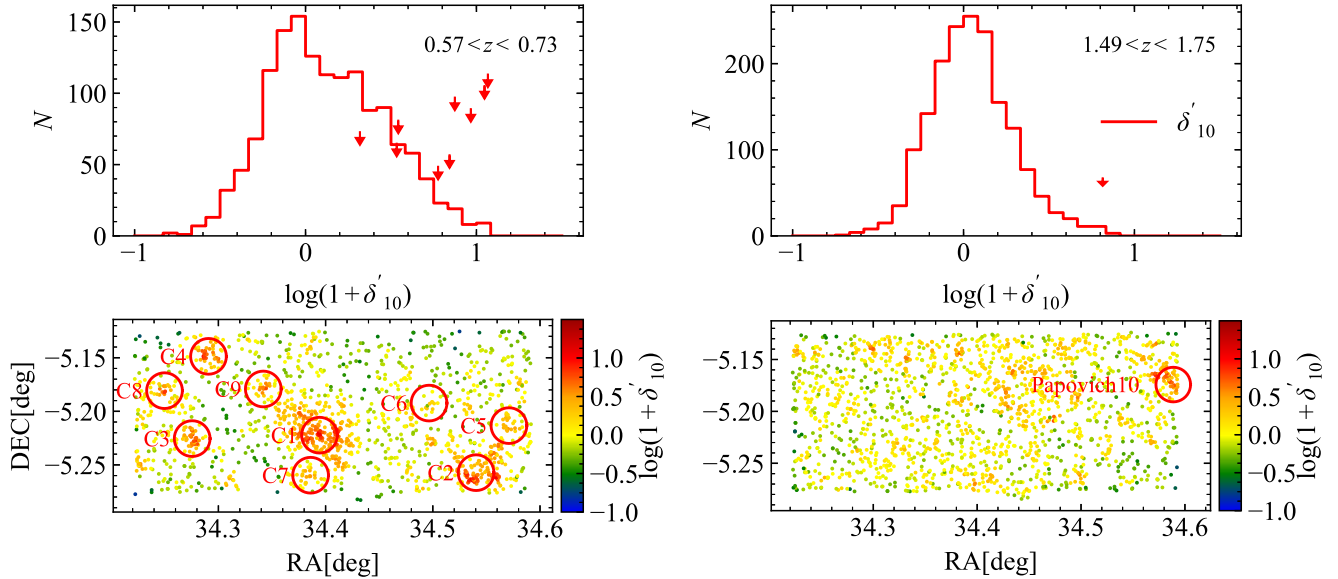}
\caption{Top: The distribution of $1+\delta'_{10}$ in the typical redshift slice. The red arrow shows the peak value of the overdensity $1+\delta'_{10}$  in these known structures. Bottom: Spatial distribution of the galaxy sample in two typical redshift slices. Points are color coded by their overdensity.
The rough locations of the known structures (groups/clusters) in the CANDELS/UDS field (Left: Cl J021734-0513 at z $\sim$ 0.65, \citealt{Galametz18}; Right: ClG 0218.3-0510 at z $\sim$ 1.62, \citealt{Papovich10}) are highlighted with red circles. Notice that the size of these circle is 1 arcmin and have nothing to do with the scale of the known structures.
}
\label{fig08}
\end{figure*}

It has been proven that our determination of overdensity, based on the Bayesian density estimator, is competent to trace the structures at high redshifts.
Recently, \cite{Galametz18} present a large-scale galaxy structure Cl J021734-0513 at $z \sim 0.65$, containing $\sim 20$ galaxy groups and clusters, which is located in the CANDELS/UDS field. The cluster candidats (`C', followed by a numerical index) are listed in \cite{Galametz18}, (e.g, C1, which is the main component of a reported large-scale structure).
The cluster candidates are well detected in Figure \ref{fig08}. Compared to the other cluster candidates, the component C6 exhibits the relative lowest signal of overdensities. Although this large scale structure is mainly composed of fainter galaxies which are excluded by the magnitude cut, its overdensity value (C6), denoted by the leftmost arrow, is still perceptible. In addtion, there is an well known structure, ClG 0218.3-0510, at z = 1.62 in the UDS fields, reported by \cite{Papovich10}. As shown in the right panel, our overdensity map  has a good performance in tracing the high-$z$ clusters.

\end{document}